\documentclass{iopart}
\usepackage[T1]{fontenc}
\usepackage{graphicx}
\usepackage{url,graphics,epsfig}
\usepackage{amssymb}

\begin{document}
\setlength{\arraycolsep}{2.5pt}             
\jl{2}
%
%
%
\def\etal{{\it et al~}}
\def\newblock{\hskip .11em plus .33em minus .07em}
%
%
%
%
%
%
\title[Photoionization Cross-Sections of halogen-like ions]{Photoionization cross section calculations for 
the halogen-like ions Kr$^+$ and Xe$^+$}

\author{B M McLaughlin$^{1,2}$\footnote[1]{Corresponding author, E-mail: b.mclaughlin@qub.ac.uk}
               and C P Ballance$^{3}$\footnote[2]{Corresponding author, E-mail: ballance@physics.auburn.edu}}

\address{$^{1}$Centre for Theoretical Atomic, Molecular and Optical Physics,
                            School of Mathematics and Physics,
                            The David Bates Building, 7 College Park, 
                            Queen's University Belfast, Belfast BT7 1NN, UK}

\address{$^{2}$Institute for Theoretical Atomic and Molecular Physics,
                            Harvard Smithsonian Center for Astrophysics, 
                            Cambridge, MA 02138, USA}
\address{$^{3}$Department of Physics,  206 Allison Laboratory,
                            Auburn University, Auburn, AL 36849, USA}

\begin{abstract}
Photoionization cross sections calculations on the halogen-like ions; Kr$^+$ and Xe$^+$ 
have been performed for a photon energy range from each ion threshold to 15 eV,
using large-scale close-coupling calculations within the  Dirac-Coulomb R-matrix approximation.  
The results from our theoretical work are compared with recent 
measurements made at the ASTRID merged-beam
set-up at the University of Aarhus in Denmark and from the Fourier transform
ion cyclotron resonance (FT-ICR) trap method 
at the SOLEIL synchrotron radiation facility in Saint-Aubin, France  \cite{bizau11} 
and the Advanced Light Soure (ALS) \cite{Mueller12,Alex12}.
For each of these complex ions our theoretical cross section results 
over the photon energy range investigated 
are seen to be in excellent agreement with experiment.
Resonance energy positions and quantum defects
 of the prominent Rydberg resonances series identified in the spectra are 
 compared with experiment for these  complex halogen like-ions.
\\
\\
\noindent
(Figures are in colour only in the online version)
\end{abstract}
%
%
%

\vspace{0.25cm} 
{\begin{flushleft}
 Short title: Photoionization cross sections for halogen-like ions\\
\vspace{0.25cm}
J. Phys. B: At. Mol. Opt. Phys: \today
\end{flushleft}}
\maketitle
%
%
%
%

%
\section{Introduction}
Most of the known matter in the Universe is in a plasma state and our
information about the Universe is carried by photons, which are dispersed and detected
for example by the orbiting Chandra X-ray Observatory. While photons travel
through stellar atmospheres and planetary nebulae, they are likely to interact with matter
and therefore with ions. This makes the study of photoionization of atoms, molecules and
their positive ions very important for astrophysicists, helping them to interpret stellar data.

Photoionization cross sections of heavy atomic elements, in low stages of ionization,
are currently of interest both experimentally and theoretically and for applications in
astrophysics. The data from such processes have many applications in planetary nebulae,
where they are of use in identifying weak emission lines of n-capture elements in NGC
3242. For example, the relative abundances of Xe and Kr can be used to determine key
physical characteristics of s-process nucleosynthesis, such as the neutron exposure
experienced by Fe-peak seed nuclei \cite{sharpee07,sterling07,sterling09,dinerstein09}. 

Xenon and Krypton ions are also of importance in man-made plasmas such as XUV light 
sources for semiconductor lithography \cite{kieft}, ion thrusters for 
space craft propulsion \cite{lerner}, and nuclear fusion plasmas
 \cite{skinner}. Xenon and Krypton ions have also been detected in cosmic objects, 
 e.g., in several planetary nebulae  and in the ejected envelopes 
 of low- and intermediate-mass stars \cite{castelli11,sterling,sterling07,sterling08,sterling09}. 
For a profound understanding of these plasmas accurate 
cross sections are required for ionization and recombination  
processes  that govern the charge balance of ions in plasmas.
Krypton and xenon ions are of particular importance in
tokamak plasmas. Injection of high-Z gases, essentially Kr and
Xe, has been proposed as a technique to mitigate disruption 
\cite{riccardo2002,nishida2002}, i.e. the uncontrolled and sudden loss of tokamak plasma
current and energy. Disruption can produce severe damage
on the vessel wall. The situation becomes more critical for
large machines such as ITER (International Thermonuclear
Experimental Reactor), where incident-loading energy may
reach several GW m$^{-2}$~\cite{federici2001,ITER}. After injection of the gases,
highly-charged Kr and Xe ions are dominant in the core of the
plasma, and low-charged and singly-charged ions are abundant
near the edge.

Over the past decade, experimentally, photoionization data on ionic targets have
been obtained using mainly two techniques: DLPP (dual
laser produced plasma) and merged beam in synchrotron
radiation facilities. While the former technique measures
photoabsorption spectra \cite{carroll86}, the latter provides absolute single
and multiple photoionization cross sections \cite{west01, kjeldsen06}. 
As indicated by Bizau and co-workers \cite{bizau11} in recent studies on Kr$^+$ and Xe$^+$ ions,
in the valence region, only the K-shell photoionization of the Kr$^+$ ion has been
reported on \cite{southworth07}. The bulk of the studies on the Xe$^+$ ion have focused 
primarily on the region of 4d inner-shell excitation and ionization using the
merged-beam technique \cite{sano96, koizumi97, andersen01, itoh01, gottwald99}. 
The recent work on Xe$^{+}$ of Bizau and co-workers \cite{bizau11}  together with the 
ongoing high resolution  studies at  the ALS \cite{Mueller12,Alex12}, in the threshold region,  
make it pertinent to have suitable theoretical results available to compare with. 
As pointed out in the recent work of Bizau and co-workers \cite{bizau11},
no studies using the above experimental techniques have been published on the
Kr$^+$ ion or on the Xe$^+$ ion in any other energy range and, in
particular, close to the thresholds, where strong resonance features dominate the respective
cross sections. Furthermore, no theoretical results on the photoionization processes are available for these
ions outside the region of 4d inner-shell excitation in the Xe$^+$ ion. In order to address 
these limitations, particularly for astrophysical applications, 
we have carried out large scale photoionization cross section calculations in the threshold 
region for both singly ionized ions of krypton and xeon. 
Where possible we have benchmarked our theoretical work with the available experimental data 
in order to provide confidence in our work for applications.

The layout of this paper is as follows. Section 2 presents a brief outline of the theoretical work. 
Section 3 details the results obtained. Section 4 presents a discussion 
and a comparison of the results obtained between experiment and theory. 
Finally in section 5 conclusions are drawn from the present investigation.

\section{Theory}

Recent modifications to the Dirac-Atomic-R-matrix-Codes 
(DARC)~\cite{venesa12} has now made it feasible to  study
photoionization of heavy complex systems of prime interest to 
astrophysics and plasma applications by including hundreds
of target levels in the close-coupling calculations.
This enables photoionization calculations  on complex ions such as singly 
ionized ions of Krypton and Xeon, the focus of the current investigation
to be carried out at the same degree of accuracy  as those for electron impact excitation. 
Such extensions to the DARC codes have allowed us recently to address the 
complex problem of  trans-iron element single photon ionization of Se$^{+}$ ions.   
In the present study we apply this suite of DARC codes to calculate detailed photoionization (PI) cross sections
on the halogen-like ions, Kr$^{+}$ and Xe$^+$.  
Recent experimental measurements \cite{bizau11} have been made on these systems but limited 
theoretical work is available to compare with.
Photoionization cross sections on these halogen-like ions are 
performed for the ground and the excited metastable levels 
associated with the $\rm ns^2np^5$ configuration 
(n=4 and 5 respectively for Kr$^+$ and  Xe$^+$ ions) in order
to benchmark our theoretical results  with recent high resolution experimental 
measurements  \cite{bizau11}. 
 
 For both the ground and metastable initial states, the outer region electron-ion collision 
problem for each ion was solved (in the resonance region below and
 between all thresholds) using a suitably chosen fine
energy mesh of 5$\times$10$^{-8}$ Rydbergs ($\approx$ 0.68 $\mu$eV) 
to fully resolve all the extremely narrow resonance structure in the appropriate photoionization cross sections. 
The jj-coupled Hamiltonian diagonal matrices were adjusted so that the theoretical term
energies matched the recommended experimental values of NIST \cite{Ralchenko2010}. We note that this energy
adjustment ensures better positioning of resonances relative to all thresholds included in
the calculation.  In the energy region considered here we can see from Tables \ref{tab1} and \ref{tab2} 
the shift to the $\rm ^1D_2$ and $\rm^1S_0$  thresholds is minimal. For this reason in
 Table \ref{tab3} - \ref{tab5} we only include experimental series limits.

In order to compare directly with the available experimental measurements for each of these ions, 
we have  convoluted the theoretical PI cross sections with a gaussian function of the appropriate full width half maximum (FWHM) 
and statistically averaged the results for the ground and metastable states.
 
\subsection{Kr$^+$}
Photoionization (PI)  cross section calculations on the Kr$^+$ 
complex were carried out retaining 326-levels 
in our close-coupling calculations with the Dirac-Atomic-R-matrix-Codes (DARC). 
In R-matrix theory, all photoionization cross section calculations require 
 the generation of atomic orbitals based primarily on 
 the atomic structure of the residual ion.
 The present theoretical work for the photoionization of the Kr$^+$ ion employs
 relativistic atomic orbitals up to n=4 generated for the residual Kr$^{2+}$ ion,
 which were calculated using the extended-optimal-level (EOL) procedure within the GRASP
 structure code \cite{dyall89,grant06,grant07}. The 1s-4s, 2p-4p and 3d-4d orbitals  were  obtained 
 from an EOL calculation on the lowest 14  levels  associated with the  $\rm 4s^24p^4$, $\rm 4s4p^5$ 
 and $\rm 4s^24p^24d^2$  configurations where the remaining four configurations;
  $\rm 4s^24p^34d$, $\rm 4s^24p^24d^2$, $\rm 4s4p^44d$ and $\rm 4p^44d^2$
  were included in the calculation. Table \ref{tab1}
  illustrates a sample of our target energy levels for the residual Kr III ion 
  compared to the NIST tabulation \cite{Ralchenko2010} for the lowest 8  levels 
  associated with the $\rm 4s^24p^4$ and $\rm 4s4p^5$ configurations.

\begin{table}
\begin{center}
\caption{Comparison of the theoretical energies using the GRASP code 
                from the 326-level model,  relative energies are in Rydbergs.
	       A sample of the first 8 levels of the Kr III ion are compared  
	       with the NIST \cite{Ralchenko2010} tabulated values\label{tab1}.}
\begin{indented}
\begin{tabular}{cccccc}
\br
Level       &  Configuration 		&  Term$^{\rm a}$           	 & Energy  		& Energy 		          & $\Delta (\%)^{\rm b}$ \\
		&					&			     		&  NIST			& GRASP 			& \\	
\mr
 1  & 4s$\rm ^2$4p$\rm ^4$ 		&  $\rm ^3$P$\rm _2$          & 0.000000 		&0.000000		&    0.0 \\  
 2  & 4s$\rm ^2$4p$\rm ^4$ 		&  $\rm ^3$P$\rm _1$          & 0.041448		&0.040837                 &    0.8 \\
 3  & 4s$\rm ^2$4p$\rm ^4$ 		&  $\rm ^3$P$\rm _0$          & 0.048415 		&0.048845	         &    0.9 \\ 
 4  & 4s$\rm ^2$4p$\rm ^4$		&  $\rm ^1$D$\rm _2$          & 0.133449	 	&0.152130		&   14.0 \\
 5  & 4s$\rm ^2$4p$\rm ^4$ 		&  $\rm ^1$S$\rm _0$          & 0.301443 		&0.318538		&    5.7 \\
 6 & 4s$\rm ^1$4p$\rm ^5$		&  $\rm ^3$P$\rm ^o_2$     & 1.0564404  		&1.230028	         &   16.4 \\
 7  & 4s$\rm ^1$4p$\rm ^5$ 		&  $\rm ^3$P$\rm ^o_1$     & 1.0878727		&1.263533		&   16.1 \\
 8  & 4s$\rm ^1$4p$\rm ^5$ 		&  $\rm ^3$P$\rm ^o_0$     & 1.1075809		&1.282909	 	&   15.8 \\
\mr
\end{tabular}
\\
$^{\rm a}$ $^{2S+1}L^{\pi}_J$\\
$^{\rm b}$ GRASP, absolute percentage difference relative to NIST values\\
\end{indented}
\end{center}
\end{table}

%
%

\begin{table}
\begin{center}
\caption{Comparison of the theoretical energies using the GRASP code 
                from the 326-level model,  relative energies are in Rydbergs.
	       A sample of the lowest 8 levels of the Xe III ion are compared  
	       with the NIST \cite{Ralchenko2010} tabulated values\label{tab2}.}
\begin{indented}
\begin{tabular}{cccccc}
\br
Level       &  Configuration 		&  Term$^{\rm a}$           	 & Energy  		& Energy 		          & $\Delta (\%)^{\rm b}$ \\
		&					&			     		&  NIST			& GRASP 			& \\	
\mr
 1  & 5s$\rm ^2$5p$\rm ^4$ 		&  $\rm ^3$P$\rm _2$          & 0.000000 		&0.000000		&    0.0 \\  
 2  & 5s$\rm ^2$5p$\rm ^4$ 		&  $\rm ^3$P$\rm _1$          & 0.074087		&0.074275                 &    0.3 \\
 3  & 5s$\rm ^2$5p$\rm ^4$ 		&  $\rm ^3$P$\rm _0$          & 0.089253 		&0.082014	         &    8.1 \\ 
 4  & 5s$\rm ^2$5p$\rm ^4$		&  $\rm ^1$D$\rm _2$          & 0.155815	 	&0.173148		&  11.1 \\
 5  & 5s$\rm ^2$5p$\rm ^4$ 		&  $\rm ^1$S$\rm _0$          & 0.328994 		&0.334735		&    1.7 \\
 6 & 5s$\rm ^1$5p$\rm ^5$		&  $\rm ^3$P$\rm ^o_2$      & 0.895434  		&0.904921	         &    2.2  \\
 7  & 5s$\rm ^1$5p$\rm ^5$ 		&  $\rm ^3$P$\rm ^o_1$      & 0.943783		&0.953095		&    1.0 \\
 8  & 5s$\rm ^1$5p$\rm ^5$ 		&  $\rm ^3$P$\rm ^o_0$      &0.987210		&0.987768	 	&    0.1 \\
\mr
\end{tabular}
\\
$^{\rm a}$ $^{2S+1}L^{\pi}_J$\\
$^{\rm b}$ GRASP,  absolute percentage difference relative to NIST values\\
\end{indented}
\end{center}
\end{table}

 Photoionization cross section calculations for 
  this complex trans-iron element included 
  all 326 levels arising from the seven configurations: 
 $\rm 4s^24p^4$, $\rm 4s4p^5$, $\rm 4s^24p^34d$, $\rm 4s^24p^24d^2$, $\rm 4p^6$,
 $\rm 4s4p^44d$ and $\rm 4p^44d^2$ in the close-coupling expansion.   
 PI cross section calculations with this 336-level model were performed in 
 the Dirac-Coulomb approximation using the DARC codes.  

The R-matrix boundary radius of 7.44 Bohr radii  was sufficient to envelop
 the radial extent of all the atomic orbitals of the residual Kr$^{2+}$ ion. A basis of 16 continuum
 orbitals was sufficient to span the incident experimental photon energy
 range from threshold  up to 40 eV. The 326-state model produced a maximum of
 1511 coupled channels in our scattering work with Hamiltonian matrices  of dimension
of the order of 24,354 by 24,354 in size. Due to dipole selection rules, 
 for total ground state photoionization we need only to consider the 
 bound-free dipole matrices, $\rm 2J^{\pi}=3^o \rightarrow 2J^{\pi}=1^e,3^e,5^e$ 
 whereas for the excited metastable states only the $\rm 2J^{\pi}=1^o \rightarrow 2J^{\pi}=1^e,3^e$ 
ones are required.
%
%

\begin{table}
\begin{center}
\caption{Principal quantum number ($\rm n$), resonance
    	       energies E (eV) and quantum defect ($\mu$) from
     	       experimental measurements of Kr$^+$  \cite{bizau11}
   	       compared  with present theoretical estimates from the QB method.
	       The Rydberg series $\rm 4s^24p^4(^1D_2)\,nd$ originating from the $\rm ^2P^o_{3/2}$ ground state and
	       the $\rm ^2P^o_{1/2}$ metastable 
	      state of Kr$^{+}$  due to $\rm 4p \rightarrow nd$ transitions are tabulated \label{tab3}.}

\begin{tabular}{cccc@{\ \ \ \ \ \ \ \ \ \ \ \ }cccc}
\hline\hline\\
\multicolumn{3}{c}{\bfseries Initial Kr$^+$ state}&&\multicolumn{3}{c}{\bfseries Initial Kr$^+$ state}\\
\multicolumn{3}{c}{$\rm \:4s^24p^5\ (^2P^{o}_{3/2})$}&&\multicolumn{3}{c}{$\rm \:4s^24p^5\ (^2P^{o}_{3/2})$}\\
\hline
\multicolumn{3}{c}{Rydberg Series}&&\multicolumn{3}{c}{Rydberg Series}\\[.02in]
\multicolumn{3}{c}{\small{$\rm \:4s^24p^4(^1D_2)\,nd $}}&&\multicolumn{3}{c}{\small{$\rm \:4s^24p^4(^1D_2)\,nd$}}\\
n			&E (eV)			&$\mu$	&	&n		&E (eV)&$\mu$\\[.02in]
\hline\\
\multicolumn{3}{c}{[Theory]}&&\multicolumn{3}{c}{[Experiment]}\\[.02in]
6	&24.579		&0.16				&			&6		&24.562 		&0.19 \\
7	&25.002		&0.19				&			&7		&24.989 		&0.23 \\
8	&25.284		&0.19				&			&8 		&25.280 		&0.20 \\
9	&25.473		&0.20				&			&9 		&25.475 		&0.19 \\
10	&25.603		&0.25				&			&10		&25.605 		&0.23 \\
11	&25.713		&0.16				&			&11		&25.710		&0.19\\
12	&25.786		&0.18				&			&12		&25.785		&0.19 \\
13	&25.844		&0.19				&			&13		&25.842 		&0.23 \\
14	&25.890		&0.19				&			&--		&-- 			&-- \\
15	&25.927		&0.19				&			&--		&-- 			&-- \\
$\cdot$		&$\cdot$					&			&		&$\cdot$	&$\cdot$				&-\\
$\infty$		&26.176$^{\dagger}$		&			&		&$\infty$	&26.176$^{\dagger}$	&\\
\hline\hline\\
\multicolumn{3}{c}{\bfseries Initial Kr$^+$ state}&&\multicolumn{3}{c}{\bfseries Initial Kr$^+$ state}\\
\multicolumn{3}{c}{$\rm \:4s^24p^5\ (^2P^{o}_{1/2})$}&&\multicolumn{3}{c}{$\rm \:4s^24p^5\ (^2P^{o}_{1/2})$}\\
\hline
\multicolumn{3}{c}{Rydberg Series}&&\multicolumn{3}{c}{Rydberg Series}\\[.02in]
\multicolumn{3}{c}{\small{$\rm \:4s^24p^4(^1D_2)\,nd $}}&&\multicolumn{3}{c}{\small{$\rm \:4s^24p^4(^1D_2)\,nd$}}\\
n			&E (eV)			&$\mu$	&	&n		&E (eV)&$\mu$\\[.02in]
\hline\\
\multicolumn{3}{c}{[Theory]}&&\multicolumn{3}{c}{[Experiment]}\\[.02in]
6	&23.914		&0.16				&		&6 		&23.902 		&0.18\\
7	&24.336		&0.19				&		&7		&24.334 		&0.20 \\
8	&24.619		&0.19				&		&8		&24.613 		&0.21 \\
9	&24.807		&0.20				&		&9 		&24.811   		&0.17 \\
10	&24.937		&0.25				&		&10 		&24.989   		&0.23 \\
11	&25.047		&0.16				&		&11		&25.047   		&0.15 \\
12	&25.121		&0.18				&		&12		&25.119  		&0.20 \\
13	&25.178		&0.18				&		&13		&25.180	   	&0.15 \\
14	&25.225		&0.19				&		&--		&--	   		&-- \\
15	&25.262		&0.19				&		&--		&--	   		&-- \\
$\cdot$		&$\cdot$					&-		&	&$\cdot$	&$\cdot$				&-\\
$\infty$		&25.510$^{\dagger}$		&		&	&$\infty$	&25.510$^{\dagger}$	&\\
\hline\hline\\
$^{\dagger}$NIST  tabulations \cite{Ralchenko2010}\\
\end{tabular}
\end{center}
\end{table}

%
%
%
%
%

%
%

\begin{table}
\begin{center}
\caption{Principal quantum number ($\rm n$), resonance
    	       energies E (eV) and quantum defect ($\mu$) from
     	       experimental measurements of Kr$^+$  \cite{bizau11}
   	       compared  with present theoretical estimates from the QB method.    	      
	       The Rydberg series $\rm 4s^24p^4(^1S_0)\,nd$ originating from the $\rm ^2P^o_{3/2}$ ground state and  $\rm ^2P^o_{1/2}$ metastable 
	      state of Kr$^{+}$  due to $\rm 4p \rightarrow nd$ transitions are tabulated\label{tab4}.}

\begin{tabular}{cccc@{\ \ \ \ \ \ \ \ \ \ \ \ }cccc}
\hline\hline\\
\multicolumn{3}{c}{\bfseries Initial Kr$^+$ state}&&\multicolumn{3}{c}{\bfseries Initial Kr$^+$ state}\\
\multicolumn{3}{c}{$\rm \:4s^24p^5\ (^2P^{o}_{3/2})$}&&\multicolumn{3}{c}{$\rm \:4s^24p^5\ (^2P^{o}_{3/2})$}\\
\hline
\multicolumn{3}{c}{Rydberg Series}&&\multicolumn{3}{c}{Rydberg Series}\\[.02in]
\multicolumn{3}{c}{\small{$\rm \:4s^24p^4(^1S_0)\,nd $}}&&\multicolumn{3}{c}{\small{$\rm \:4s^24p^4(^1S_0)\,nd$}}\\
n			&E (eV)			&$\mu$	&	&n		&E (eV)		&$\mu$\\[.02in]
\hline\\
\multicolumn{3}{c}{[Theory]}&&\multicolumn{3}{c}{[Experiment]}\\[.02in]
4		&24.520				&0.28			&		&4		&24.501 				&0.29 \\
5		&25.954				&0.34			&		&5		&25.910 				&0.38 \\
6		&26.761				&0.34			&		&6 		&26.740 				&0.38 \\
7		&27.235				&0.34			&		&7  		&27.230 				&0.35\\
8		&27.534				&0.34			&		&8		&27.530				& 0.38  \\
9		&27.736				&0.34			&		&9		&--					&--    \\
10		&27.878				&0.34			&		&10		&--					&--  \\
11		&27.983				&0.33			&		&11		&--					&--  \\
12		&28.061				&0.33			&		&12		&--					&--  \\
$\cdot$	&$\cdot$				&				&		&$\cdot$	&$\cdot$				&-\\
$\infty$	&28.461$^{\dagger}$	&				&		&$\infty$  &28.461$^{\dagger}$	&\\
\hline\hline\\
\multicolumn{3}{c}{\bfseries Initial Kr$^+$ state}&&\multicolumn{3}{c}{\bfseries Initial Kr$^+$ state}\\
\multicolumn{3}{c}{$\rm \:4s^24p^5\ (^2P^{o}_{1/2})$}&&\multicolumn{3}{c}{$\rm \:4s^24p^5\ (^2P^{o}_{1/2})$}\\
\hline
\multicolumn{3}{c}{Rydberg Series}&&\multicolumn{3}{c}{Rydberg Series}\\[.02in]
\multicolumn{3}{c}{\small{$\rm \:4s^24p^4(^1S_0)\,nd $}}&&\multicolumn{3}{c}{\small{$\rm \:4s^24p^4(^1S_0)\,nd$}}\\
n			&E (eV)	&$\mu$	&	&n		&E (eV)			&$\mu$\\[.02in]
\hline\\
\multicolumn{3}{c}{[Theory]}&&\multicolumn{3}{c}{[Experiment]}\\[.02in]
4		&23.854				&0.28	&		&4		&23.845				&0.29 \\
5		&25.288				&0.34	&		&5		&25.280 				&0.20 \\
6		&26.096				&0.34	&		&6 		&26.100 				&0.33\\
7		&26.569				&0.34	&		&7  		&26.580				&0.31\\
8		&26.868				&0.34	&		&8		&--					&--     \\
9		&27.070				&0.34	&		&9		&--					&--    \\
10		&27.213				&0.34	&		&10		&--                       		&--   \\
11		&27.317				&0.33	&		&11		&--                       		&--   \\
12		&27.395				&0.33	&		&12		&--                       		&--   \\
$\cdot$	&$\cdot$				&-		&		&$\cdot$	&$\cdot$				&-\\
$\infty$	&27.795$^{\dagger}$	&		&		&$\infty$	&27.795$^{\dagger}$	&\\
\hline\hline\\
$^{\dagger}$NIST  tabulations \cite{Ralchenko2010}\\
\end{tabular}
\end{center}
\end{table}
%
%
%
\begin{figure}
\begin{center}
\includegraphics[width=\textwidth]{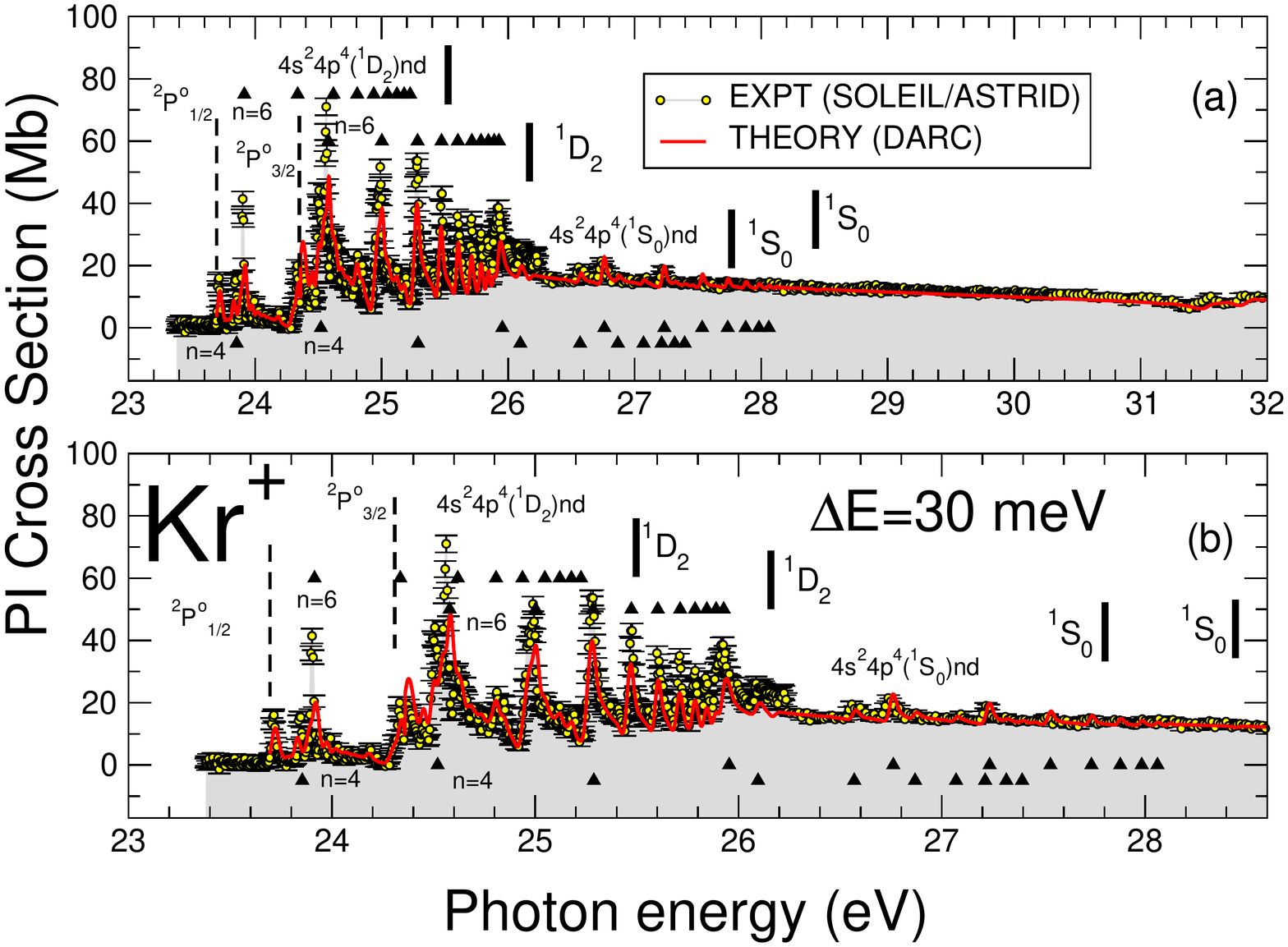}
\caption{\label{fig:theooverview}  Theoretical cross sections for photoionization (PI) of Halogen-like Kr$^\mathrm{+}$ ions.
                                            (a) The 326-state $jj$-coupling R-matrix calculations
                                            were carried out for  ground-state and metastable-state parent ions 
                                            using the recently developed DARC suite of codes. 
                                            The theoretical cross sections (solid line) were convoluted at 30 meV FWHM and
                                            statistically averaged in order to compare directly with the recent
                                            experimental results \cite{bizau11} for the energy range threhold. The prominent Rydberg 
                                            resonance series $\rm 4s^24p^4 (^1D_2, ^1S_0)nd$  converging to the 
                                            Kr$^\mathrm{2+}$ ($\rm ^1D_2$ and $\rm ^1S_0$) thresholds (b)  are tabulated in Table 1 and 2. }
\end{center}
\end{figure}

\begin{figure}
\begin{center}
\includegraphics[width=\textwidth]{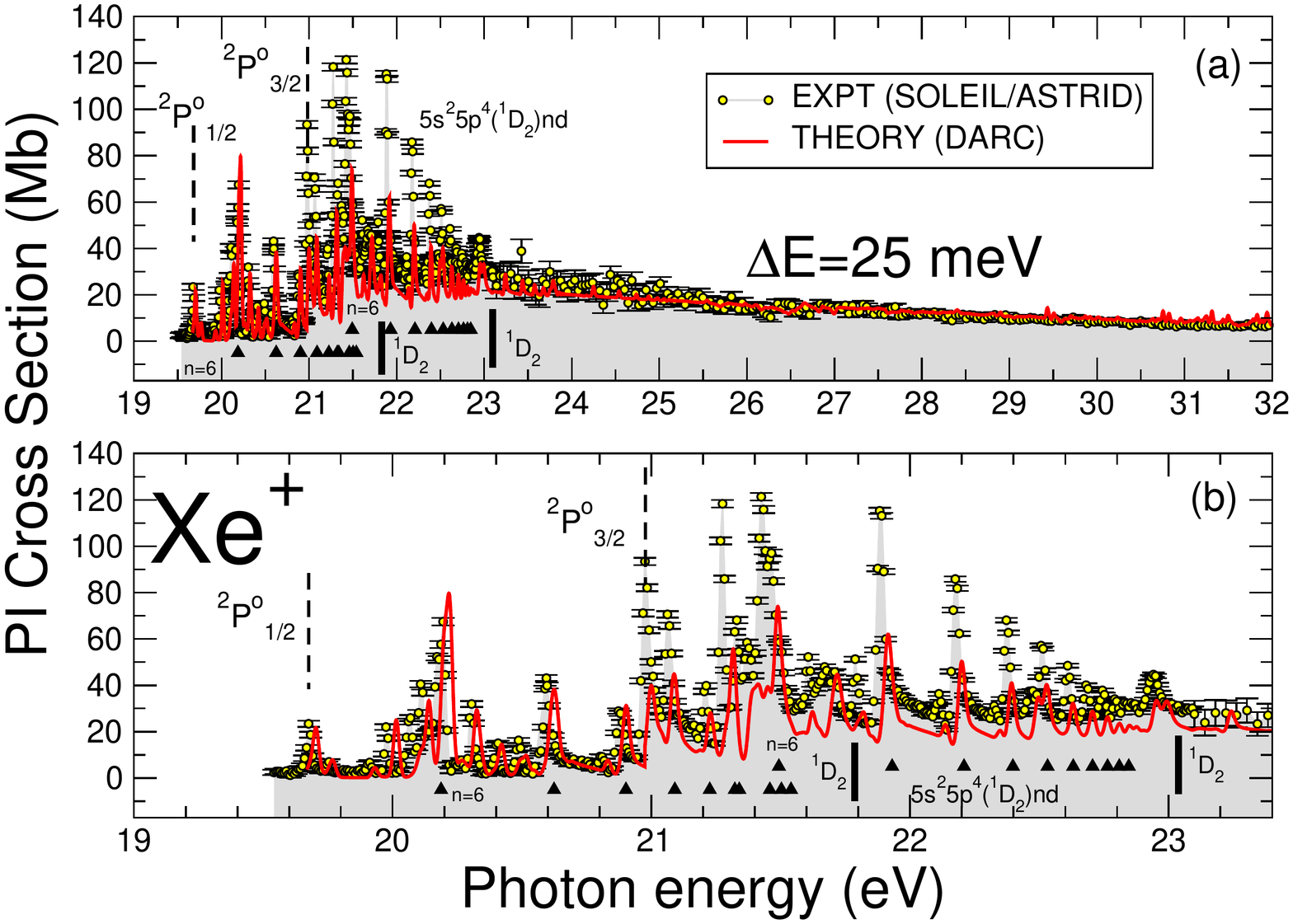}
\caption{\label{fig:theooverview1}  Theoretical cross sections for photoionization (PI) of Halogen-like Xe$^\mathrm{+}$ ions.
                                            (a)The 326-state $jj$-coupling R-matrix calculations
                                            were carried out for ground-state and metastable-state parent ions 
                                            using the recently developed DARC suite of codes. 
                                            The theoretical cross sections (solid line) were convoluted at 25 meV FWHM and
                                            statistically averaged.  The experiment data \cite{bizau11} was normalized to theory
                                             at 28 eV in order to have a direct comparison. The prominent Rydberg 
                                            resonance series $\rm 5s^25p^4 (^1D_2)nd$ converging 
                                            to the Xe$^\mathrm{2+}$ ($\rm^1D_2$) threshold (b) are tabulated in Table \ref{tab5}. }
\end{center}
\end{figure}

\subsection{Xe$^+$}
Similarly for photoionization cross section calculations on the Xe$^+$ 
system we retained 326 levels of the residual Xe$^{2+}$  ion 
in our close-coupling calculations performed with the Dirac-Atomic-R-matrix-Codes (DARC). 
Analogous PI calculations on the Kr$^+$ case, were made similarly for the Xe$^+$ ion. 
For the Xe$^+$ case we have employed relativistic n=5 atomic orbitals 
generated for the residual Xe$^{2+}$ ion,
 which were obtained using the energy-average-level (EAL) 
 procedure within the GRASP structure code on the fourteen lowest levels associated 
 with the  $\rm 5s^25p^4$, $\rm 5s5p^5$ and $\rm 5s^25p^35d^2$ configurations. Table \ref{tab2} 
 gives a sample of our results for the Xe III ion for the energies of the lowest eight levels associated 
 with the  $\rm 5s^25p^4$ and $\rm 5s5p^5$ configurations compared with 
 the NIST \cite{Ralchenko2010} tabulations.
Here again for the photoionization cross section calculations on 
  this complex trans-iron system we included 
  all 326 levels arising from the seven configurations: 
 $\rm 5s^25p^4$, $\rm 5s5p^5$, $\rm 5s^25p^35d$, $\rm 5s^25p^25d^2$, $\rm 5p^6$,
 $\rm 5s5p^45d$ and $\rm 5p^45d^2$ in the close-coupling expansion.   
 PI cross section calculations with this 326-level model were then carried out in 
 the Dirac-Coulomb approximation using the DARC codes for photon energies up 
 to about 15 eV above the ion threshold.

%
%
%
%
%
%
%

\begin{table}
\begin{center}
\caption{Principal quantum number ($\rm n$), resonance
    	       energies E (eV) and quantum defect ($\mu$) from
     	       experimental measurements of Xe$^+$  \cite{bizau11}
   	       compared  with present theoretical estimates from the QB method.
	       The Rydberg series $\rm 5s^25p^4(^1D_2)\,nd$ originating from the $\rm ^2P^o_{3/2}$ ground state and
	       the $\rm ^2P^o_{1/2}$ metastable 
	      state of Xe$^{+}$  due to $\rm 5p \rightarrow nd$ transitions are tabulated\label{tab5}.}

\begin{tabular}{cccc@{\ \ \ \ \ \ \ \ \ \ \ \ }cccc}
\hline\hline\\
\multicolumn{3}{c}{\bfseries Initial Xe$^+$ state}&&\multicolumn{3}{c}{\bfseries Initial Xe$^+$ state}\\
\multicolumn{3}{c}{$\rm \:5s^25p^5\ (^2P^{o}_{3/2})$}&&\multicolumn{3}{c}{$\rm \:5s^25p^5\ (^2P^{o}_{3/2})$}\\
\hline
\multicolumn{3}{c}{Rydberg Series}&&\multicolumn{3}{c}{Rydberg Series}\\[.02in]
\multicolumn{3}{c}{\small{$\rm \:5s^25p^4(^1D_2)\,nd $}}&&\multicolumn{3}{c}{\small{$\rm \:5s^25p^4(^1D_2)\,nd$}}\\
n			&E (eV)			&$\mu$	&	&n		&E (eV)&$\mu$\\[.02in]
\hline\\
\multicolumn{3}{c}{[Theory]}&&\multicolumn{3}{c}{[Experiment]}\\[.02in]
6	&21.493		&0.17	&		&6		&21.425 		&0.29\\
7	&21.931		&0.16	&		&7		&21.886 		&0.29\\
8	&22.209		&0.16	&		&8 		&22.177 		&0.30\\
9	&22.398		&0.17	&		&9 		&22.375 		&0.31\\
10	&22.532		&0.17	&		&10		&22.510 		&0.35 \\
11	&22.631		&0.17	&		&11		&22.610 		&0.41\\
12	&22.706		&0.17	&		&12		&22.683 		&0.51\\
13	&22.764		&0.18	&		&13		&22.747 		&0.49\\
14	&22.809		&0.20	&		&--		&--	 		&--\\
15	&22.846		&0.20	&		&--		&-- 			&--\\
$\cdot$		&$\cdot$		&		&--	&$\cdot$	&$\cdot$	&--\\
$\infty$		&23.095$^{\dagger}$	&		&	&$\infty$	&23.095$^{\dagger}$	&\\
\hline\hline\\
\multicolumn{3}{c}{\bfseries Initial Xe$^+$ state}&&\multicolumn{3}{c}{\bfseries Initial Xe$^+$ state}\\
\multicolumn{3}{c}{$\rm \:5s^25p^5\ (^2P^{o}_{1/2})$}&&\multicolumn{3}{c}{$\rm \:5s^25p^5\ (^2P^{o}_{1/2})$}\\
\hline
\multicolumn{3}{c}{Rydberg Series}&&\multicolumn{3}{c}{Rydberg Series}\\[.02in]
\multicolumn{3}{c}{\small{$\rm \:5s^25p^4(^1D_2)\,nd $}}&&\multicolumn{3}{c}{\small{$\rm \:5s^25p^4(^1D_2)\,nd$}}\\
n			&E (eV)			&$\mu$	&	&n		&E (eV)&$\mu$\\[.02in]
\hline\\
\multicolumn{3}{c}{[Theory]}&&\multicolumn{3}{c}{[Experiment]}\\[.02in]
6	&20.187	&0.17	&		&6 		&20.195 			&0.16\\
7	&20.624	&0.16	&		&7		&20.594 			&0.25\\
8	&20.902	&0.16	&		&8		&20.876 			&0.28\\
9	&21.091	&0.16	&		&9 		&21.062 			&0.34\\
10	&21.226	&0.17	&		&10 		&21.210			&0.30\\
11	&21.323	&0.17	&		&-- 		&--				&--\\
12	&21.340	&0.17	&		&--		&--				&--\\
13	&21.458	&0.18	&		&--		&--				&--\\
14	&21.503	&0.20	&		&-- 		&--				&--\\
15	&21.540	&0.20	&		&-- 		&--				&--\\
$\cdot$		&$\cdot$	&		&		&$\cdot$	&$\cdot$	&-\\
$\infty$		&21.789$^{\dagger}$	&		&	&$\infty$	&21.789$^{\dagger}$	&\\
\hline\hline\\
$^{\dagger}$NIST  tabulations \cite{Ralchenko2010}\\
\end{tabular}
\end{center}
\end{table}

\section{Results}    
Figures 1 and 2 presents our theoretical results from the 326-level model using the DARC  suite of codes.
In order to compare directly with the experimental results of Bizau and co-workers \cite{bizau11} 
we have statistically averaged the results from the ground and metastable levels and convoluted 
respective cross sections  with a Gaussian of 30 meV FWHM for Kr$^\mathrm{+}$ ions and 25 meV for Xe$^\mathrm{+}$ ions. 
In the near threshold region (cf figures 1 b and 2 b) we see clearly 
(from a comparison of the theoretical and experimental results of Bizau et al \cite{bizau11}) that 
the rich and complex resonance structure is better reproduced by the DARC calculations for Kr$^+$ than for Xe$^+$,  
which could be due to the limited resolution in those experiments.
Figure 3 shows a comparison with recent ALS measurements for Xe$^{+}$ ions in the near threshold region 
taken at the Advanced Light Soure (ALS) at a extremely high resolution of 4 meV \cite{Mueller12}.  Here we see that the 
experimental PI cross sections (apart from the region of the n=14 member of the 5s$^2$5p$^4$ ($^3$P$_1$) nd  series) 
are reproduced by our DARC calculations.

The multi-channel R-matrix eigenphase derivative (QB) technique (applicable to atomic and molecular complexes) of Berrington and 
co-workers \cite{keith1996,keith1998,keith1999} modified to cater for jj-coupling 
was used to determine the resonance parameters of the prominent series. 
The resonance width $\Gamma$ was determined from 
the inverse of the energy derivative of the eigenphase sum $\delta$ at the resonance energy $E_r$ via
\begin{equation}
\Gamma = 2\left[{\frac{d\delta}{dE}}\right]^{-1}_{E=E_r}.
\end{equation}
The results for all the resonance parameters determined from the QB method 
are presented in Tables \ref{tab3} and \ref{tab4} for Kr$^\mathrm{+}$  ions
and compared with the available experimentally determined ones. Table \ref{tab5} give similar results 
 for the corresponding Xe$^\mathrm{+}$ ion. 

\section{Discussion}
The hypothesis of statistical population is very reasonable since the
life time is huge (0.34 seconds for the case of Kr$^\mathrm{+}$) if compared to the beam transport
times. Note also that the excited $\rm ^2P^o_{1/2}$ metastable Kr$^\mathrm{+}$ ions produced
either in an ion source or ion trap rarely collide with surfaces or residual
gas before photoionization takes place. As a result, statistical
population of the ions seems consistent with their measurements both using an
ion source and an ion trap without delay\cite{bizau11}.  For the case of Kr$^\mathrm{+}$ ions studied here resonance 
positions and quantum defects are in excellent agreement with the available experimental measurements 
as can be seen from Tables \ref{tab3} and \ref{tab4}. The magnitude of the absolute cross sections are also in excellent agreement 
with experimental values as illustrated in figure 1.
 
For Xe$^\mathrm{+}$ ions the difference in the resolution between the Bizau and co-workers \cite{bizau11} (25 meV)
and current Advanced Light Source (ALS) measurements (2.2 meV) \cite{Mueller12} and that taken 
at 4 meV shown in figure 3 \cite{Alex12} is over a factor of ten better. 
Bizau and co-workers  noted that, given their
poor resolution, they do not attempt to obtain more detailed
spectroscopic information: ÒConsidering the moderate energy resolution
chosen in this work to compensate for the low density of target ions, we
have not attempted a detailed identification of the observed structures,
which are most of the time composed of several unresolved lines."  In the 
near threshold region we do not fully reproduce the resonance strengths in the experimental data 
of Bizau and co-workers \cite{bizau11} which could be due to the limited 
resolution in those experiments.
We also note here there is a discrepancy in the quantum defects for the Rydberg 
series in Xe$^\mathrm{+}$ as given in Table \ref{tab5} with the experimental values of Bizau and co-workers \cite{bizau11}.
The recent high resolution measurements (4 meV) see figure 3  made at the ALS 
by Muller and co-workers \cite{Alex12}  have made possible 
spectroscopic studies of the Xe$^\mathrm{+}$ PI spectrum in the near threshold region producing an average quantum defect
for the  $\rm 5s^25p^4(^3P_1)\,nd$  Rydberg series of 0.16 which are comparable values to 
our DARC estimates for the $\rm 5s^25p^4(^1D_2)\,nd$ series  in Table 5 \cite{Mueller12,Alex12}. 
For Xe$^+$ ions a more stringent test with recent extremely 
high resolution  experimental measurements made at the ALS at 4 meV  indicate 
(in the near threshold region, apart from the n=14 member of the Rydberg series) 
very good agreement giving us confidence in our theoretical results.

\begin{figure}
\begin{center}
\includegraphics[width=\textwidth]{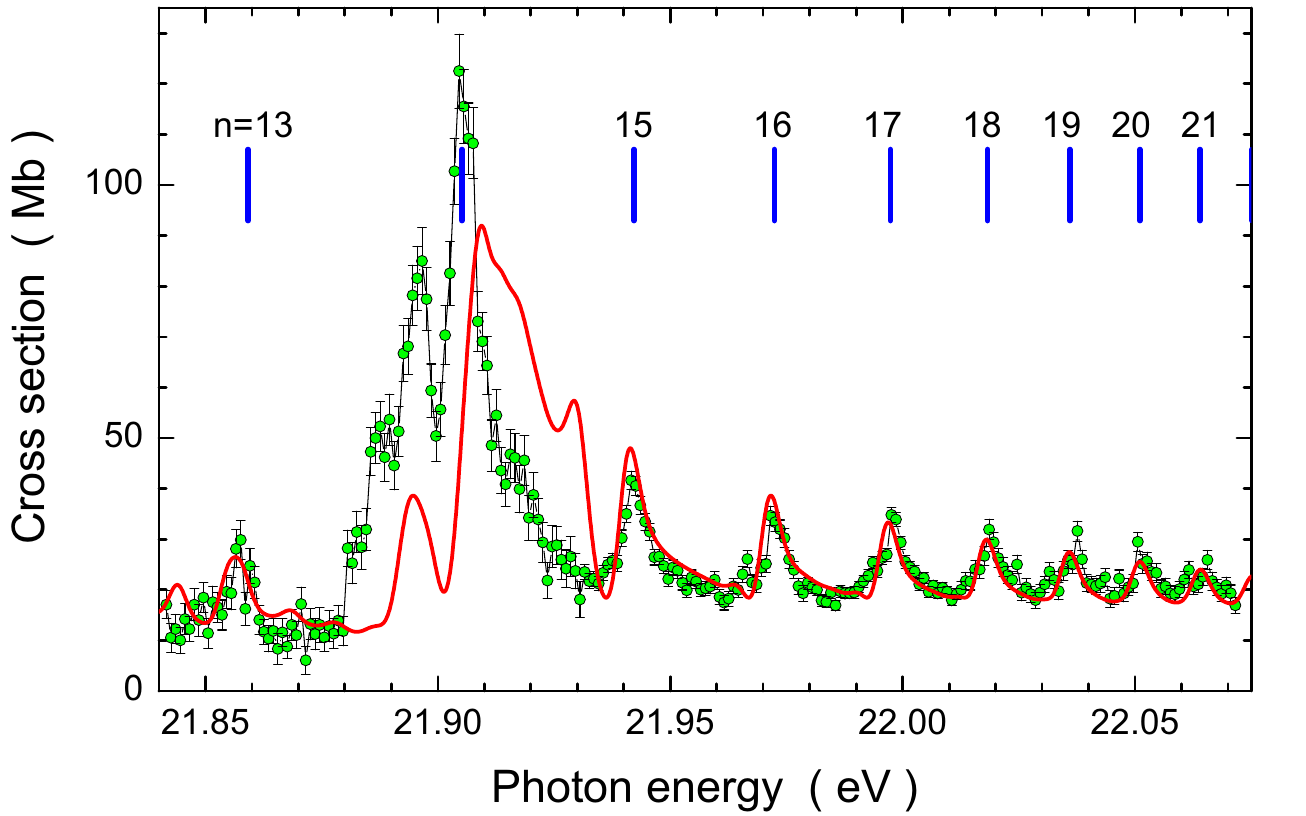}
\caption{\label{fig:4meV} Xe$^+$ ALS experimental PI cross section data (green circles) 
                 for photon energies ranging from 21.84 eV - 22.08 eV at
                a photon energy resolution of 4 meV. Results are compared 
                with theoretical results from a 326-level Dirac-Coulomb R-matrix calculation
               (red line) convoluted with a FWHM Gaussian of 4 meV and statistically 
                averaged over the ground and metastable states to simulate the experimental measurements.
                The bars mark the energies of  the [5s$^2$5p$^4$ ($^3$P$_1$) nd] resonances 
                obtained with a quantum defect of 0.16. \cite{Mueller12,Alex12} } 
 \end{center}
\end{figure}

\section{Conclusions}
State-of-the-art theoretical methods were used to investigate 
photons interacting with the halogen-like ions, Kr$^\mathrm{+}$ and Xe$^\mathrm{+}$,  
in the energy region extending to about 15 eV beyond the ionization threshold.  
Given the complexity of these halogen-like ions, throughout
 the energy region investigated the agreement  (on the photon-energy 
 scale and on the absolute PI cross-section scale)
 is better for Kr$^+$ ions than for Xe$^+$ ions with the recent experimental measurements 
of Bizau and co-workers \cite{bizau11} and our DARC calculations. However our DARC calculations 
in the near threshold region are seen to reproduce most of the recent  extremely high resolution 
ALS measurements on Xe$^+$ ions \cite{Mueller12,Alex12}.
It is seen that the photoionization cross section exhibits a wealth of resonances 
which theory is able to mostly reproduce. 
The prominent members of the Rydberg series are 
analyzed and compared with experiment.
We point out that the strength of the present study is the benchmarking 
of our theoretical work with the available experimental data. 
Such detailed comparison between theory and experiment 
strengthens the validity of our results giving confidence 
in their use for astrophysical applications. 

The photoionization cross-sections from the present study are suitable to be included into
state-of-the-art photoionization modelling codes Cloudy
and XSTAR \cite{ferland98,kallman01}  that are used to numerically
simulate the thermal and ionization structure of ionized astrophysical nebulae.

\ack
C P Ballance was supported by US Department of Energy (DoE)
grants  through Auburn University.
B M McLaughlin acknowledges support by the US
National Science Foundation through a grant to ITAMP
at the Harvard-Smithsonian Center for Astrophysics.  
We thank Dr Jean-Marc Bizau for providing us 
with the recent ASTRID/SOLEIL experimental data on these 
ions and Professor Alfred Mueller for the ALS experimental data.
The computational work was carried out at the National Energy Research Scientific
Computing Center in Oakland, CA, USA and on the Tera-grid at
the National Institute for Computational Sciences (NICS) in Knoxville, TN, USA. 
%
%
%
%
\bibliographystyle{iopart-num}
\bibliography{halogen}
\end{document}